\documentclass[aps,twocolumn,showpacs,preprintnumbers,amsmath,amssymb,prl,10pt,superscriptaddress,ltxgridinfo,floatfix]{revtex4}
\usepackage{graphicx,amsbsy, color}

\begin{document}

\title{Charge-orbital-lattice coupling effects in the $dd$-excitation profile of one dimensional cuprates}

\date{\today}

\author{J.J. Lee}
\affiliation{Stanford Institute for Materials and Energy Sciences, SLAC National Accelerator Laboratory, Menlo Park, CA 94025, USA}
\affiliation{Departments of Physics and Applied Physics, and Geballe Laboratory for Advanced Materials, Stanford University, Stanford, CA, 94305, USA}
\author{B. Moritz}
\affiliation{Stanford Institute for Materials and Energy Sciences, SLAC National Accelerator Laboratory, Menlo Park, CA 94025, USA}
\affiliation{Department of Physics and Astrophysics, University of North Dakota, Grand Forks, ND 58202, USA}
\affiliation{Department of Physics, Northern Illinois University, DeKalb, IL 60115, USA}
\author{W.S. Lee}
\author{M. Yi}
\affiliation{Stanford Institute for Materials and Energy Sciences, SLAC National Accelerator Laboratory, Menlo Park, CA 94025, USA}
\affiliation{Departments of Physics and Applied Physics, and Geballe Laboratory for Advanced Materials, Stanford University, Stanford, CA, 94305, USA}

\author{C.J. Jia}
\author{A. P. Sorini}
\affiliation{Stanford Institute for Materials and Energy Sciences, SLAC National Accelerator Laboratory, Menlo Park, CA 94025, USA}

\author{K. Kudo}
\affiliation{Department of Physics, Okayama University, Okayama 700-8530, Japan}
\author{Y. Koike}
\affiliation{Department of Applied Physics, Tohoku University, Sendai, Japan}

\author{K.J. Zhou}
\affiliation{SLS, Paul Scherrer Institute, CH-5232, Villigen PSI, Switzerland}
\author{C. Monney}
\affiliation{Fritz-Haber-Institut der Max Planck Gesellschaft, Faradayweg 4-6, D-14195 Berlin, Germany}
\affiliation{SLS, Paul Scherrer Institute, CH-5232, Villigen PSI, Switzerland}
\author{V. Strocov}
\author{L. Patthey}
\author{T. Schmitt}
\affiliation{SLS, Paul Scherrer Institute, CH-5232, Villigen PSI, Switzerland}

\author{T. P. Devereaux}
\author{Z.X. Shen}
\affiliation{Departments of Physics and Applied Physics, and Geballe Laboratory for Advanced Materials, Stanford University, Stanford, CA, 94305, USA}
\affiliation{Stanford Institute for Materials and Energy Sciences, SLAC National Accelerator Laboratory, Menlo Park, CA 94025, USA}

\begin{abstract}
We identify $dd$-excitations in the quasi-one dimensional compound Ca$_2$Y$_2$Cu$_5$O$_{10}$ using resonant inelastic x-ray scattering. By tuning across the Cu $L_{\mathrm{3}}$-edge, we observe abrupt shifts in the $dd$-peak positions as a function of incident photon energy. This observation demonstrates orbital-specific coupling of the high-energy excited states of the system to the low-energy degrees of freedom.  A Franck-Condon treatment of electron-lattice coupling, consistent with other measurements in this compound, reproduces these shifts, explains the Gaussian lineshapes, and highlights charge-orbital-lattice renormalization in the high energy $d$-manifold. 
\end{abstract}
 
\pacs{78.70.Ck,71.38.-k,74.72.-h,7b1.70.Ch}

\maketitle
Novel quantum phases in solids often emerge via coupling between the charge, spin, lattice, and orbital degrees of freedom. To gain insight into the formation of these novel phases, it is important to determine the energy scales of the collective excitations associated with these degrees of freedom and to understand their mutual couplings. Spin and lattice excitations have traditionally been studied through scattering experiments including Raman\cite{RevModPhys.79.175}, neutron\cite{JPSJ.75.111003, JPSJ.81.011007, PhysRevB.84.085132}, and inelastic x-ray scattering\cite{0034-4885-63-2-203,RevModPhys.79.175}. Recently, resonant inelastic x-ray scattering (RIXS) has emerged as a powerful tool to study various collective excitations\cite{RevModPhys.83.705, RevModPhys.73.203}. In particular, it has been demonstrated that RIXS can probe the interaction between distinct degrees of freedom, such as electron-phonon\cite{PhysRevLett.110.265502} and spin-orbital\cite{ISI:000303451900038} coupling in materials. 

RIXS at the transition metal $L$-edge is especially suited to probing the excitations within the $d$-orbital manifold
\cite{PhysRevB.54.4405,PhysRevLett.92.117406}. These so-called $dd$-excitations  cannot be excited through direct optical transitions due to dipole selection rules, and are difficult to measure with neutron scattering\cite{PhysRevB.84.085132}. Their energies are largely determined by the oxygen configuration through hybridization and crystal field splitting; therefore $dd$-excitations are, in principle, sensitive to lattice distortions, such as Jahn-Teller and phonon modes, and contain rich information about electron-lattice-orbital coupling in transition metal oxide systems. However in the past, the coupling of $dd$-excitations to lattice degrees of freedom has not been demonstrated due to the insufficient energy resolution of RIXS measurements.

\begin{figure}[t]
\includegraphics[scale=.5]{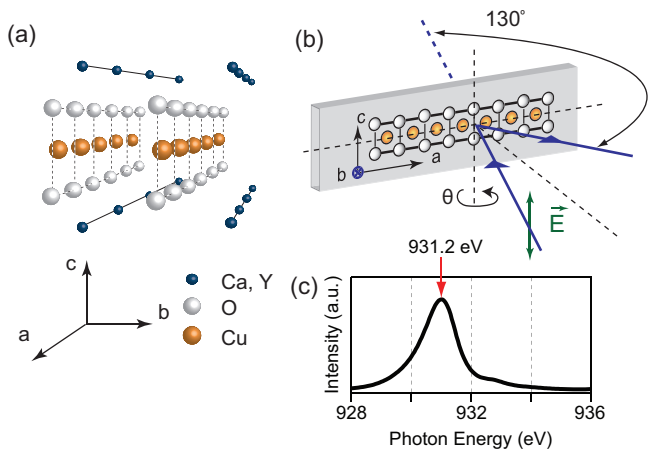}
\caption{(Color online) (a) A sketch of the crystal structure of Ca$_{2}$Y$_{2}$Cu$_{5}$O$_{10}$.  (b) Scattering geometry relative to the CuO$_{2}$ chains. The blue arrows indicate the incident and out-going x-rays. The green arrow indicates the electric field (i.e.polarization) of the incident x-rays. (c) XAS (TFY) of the undoped compound, with the red arrow indicating the copper $L_{3}$-edge. \label{fig1}}
\end{figure}

Here we present a Cu $L_{\mathrm{3}}$-edge RIXS measurement of the $dd$-excitation profile for the quasi-one-dimensional edge-sharing cuprate Ca$_{2}$Y$_{2}$Cu$_5$O$_{10}$ (CYCO) \cite{KudoPRB2005,Kabasawa200565,MatsudaPRB2005}. CYCO consists of edge-sharing Cu and O plaquette chains running parallel to the $a$-axis (see Fig.~\ref{fig1}(a)) with interstitial Ca and Y atoms providing a charge reservoir. The magnetic superexchange energy $J$ of the system is expected to be much smaller \cite{MizunoPRB1998} than those of corner-sharing quasi-one-dimensional cuprates such as Sr$_2$CuO$_3$ \cite{MizunoPRBR1998} according to the Goodenough-Kanamori-Anderson rules\cite{GoodenoughJPCS1958,KanamoriJPCS1959} and the near-perpendicular Cu-O-Cu bond angles in CYCO . Indeed, inelastic neutron scattering studies have found a ferromagnetic intra-chain exchange $J{\sim}7$-$15$ meV\cite{MatsudaPRB2001,MatsudaPRB2005,KuzianPRL2012}.  Since the magnetic energy scale is much smaller than the resolution of our measurements ($\sim$140 meV), the observations reported in this letter are unlikely to be associated with magnetic degrees of freedom.

We identify the energy scales and respective orbital characters of the $dd$-excitations in CYCO through the scattering geometry dependence of the RIXS cross-section. A RIXS map of energy loss versus incident photon energy reveals an abrupt shift in the excitation peak energy as one scans across the Cu $L_{\mathrm{3}}$-edge, indicating a significant coupling between the charge, orbital, and lattice degrees of freedom. A Franck-Condon (FC) treatment for the coupling between the electronic and lattice degrees of freedom reproduces the observed structures, consistent with an analysis conducted at the O $K$-edge for the same family\cite{PhysRevLett.110.265502}. Our results demonstrate how high energy orbital excitations can be significantly renormalized by the lattice degrees of freedom, which have a much lower energy scale.

Experiments were performed at the ADRESS beamline of the Swiss Light Source at the Paul Scherrer Institut using the SAXES spectrometer\cite{ghiringhelli:113108,StrocovADRESS}. Fig. ~\ref{fig1}(b) shows a schematic of the scattering geometry in relation to the crystal axes and incident polarization, fixed normal to the scattering plane ($\sigma$ polarization). In our setup there is no polarization discrimination of the scattered photons at the detector.  RIXS measurements were performed with a fixed scattering angle of 130$^{\circ}$. The sample was rotated by an angle $\theta$ to access different polarization configurations, allowing us to identify the character of the $dd$-excitations. The energy resolution was set to 140 meV.  The sample temperature was held at 30 K, which is above the magnetic transition temperature of the material\cite{KudoPRB2005}. Fig.  ~\ref{fig1}(c) shows the x-ray absorption spectrum (XAS) of CYCO, which was measured with total fluorescence yield (TFY), with the arrow indicating the incident photon energy corresponding to the Cu $L_{\mathrm{3}}$-edge (931.2 eV).

\begin{figure}[b]
\includegraphics[scale = .5]{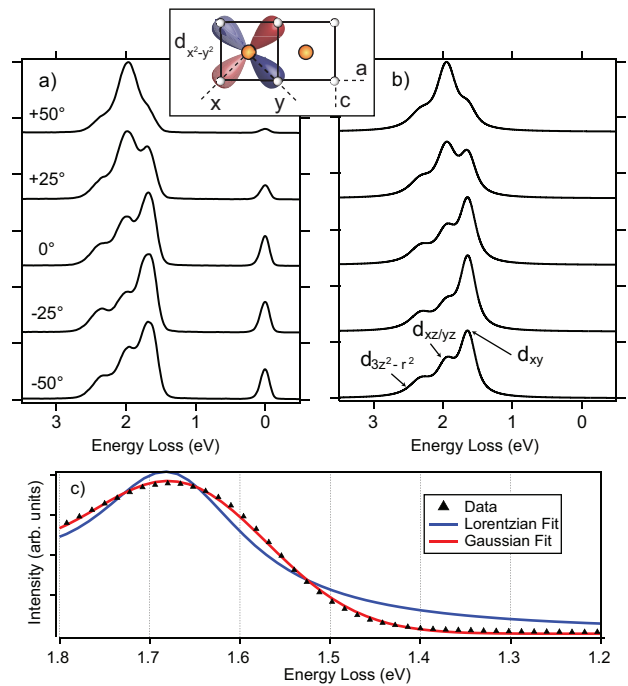}
\caption{(a) Angular dependence of the $dd$-excitation spectra at the Cu $L_{\mathrm{3}}$-edge resonance. (b) Calculated $dd$-excitation intensities for each given scattering geometry. The characters of each $dd$-peak are given in the figure. The inset gives the local coordinate convention for the $d$-orbitals with respect to the crystal axes. (c) Comparison of fits with different lineshapes of the spectra taken at 0 degrees. We find that a Gaussian profile provides a much better fit relative to the Lorentzian. \label{fig2}}
\end{figure}

Fig. ~\ref{fig2}(a) shows RIXS spectra taken at the absorption edge for various incident angles $\theta$, relative to specular reflection. Energy loss, defined as the difference between the incident and scattered photon energies, measures the energy of the various excitations created during the RIXS process. The three peaks seen in the range of 1.5 to 2.5 eV correspond to the $dd$-excitations and show little dispersion as a function of $\theta$, indicative of their localized nature. Intriguingly, the intensities of the $dd$-excitations vary dramatically as a function of angle. Such an intensity modulation arises from the change in scattering matrix elements due to changes in photon polarization with respect to the orbital spatial symmetry. Therefore, the orbital nature of the $dd$-excitations can be determined by modelling this intensity modulation, as has been demonstrated in other cuprates\cite{PhysRevB.77.224107,ddNJP}.

\begin{figure*}[!t]
\includegraphics[scale = .55]{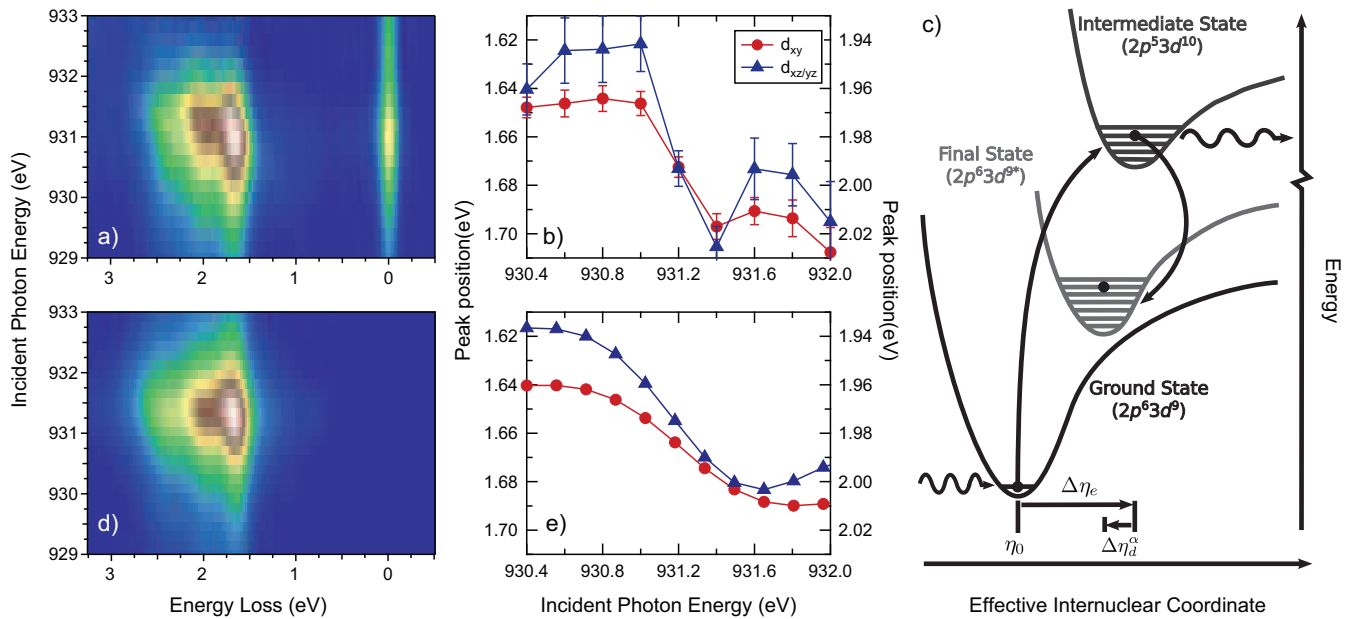}
\caption{(a) RIXS energy-loss map for CYCO at specular reflection. A visible shift in the peaks can be seen as one tunes across the Cu $L_{\mathrm{3}}$-edge resonance. (b) Plot of the peak position fit for the $d_{xy}$ and $d_{xz/yz}$ excitations. The peaks shift on the order of 50 - 80 meV.  (c) Schematic showing the Franck-Condon process at the Cu$L_{\mathrm{3}}$-edge. (d) Calculated RIXS map using a Franck-Condon treatment for the lattice coupling to the atomic-level model for the electronic and orbital degrees of freedom. (e) Fits of the peak positions from the calculated RIXS map showing a similar shift over an equivalent incident energy window.  \label{fig4}}
\end{figure*}

Following the procedures described in Ref. ~\onlinecite{ddNJP}, we use an atomic model to calculate the angular dependence of the $dd$-excitations, as shown in Fig. 2 (b), neglecting the elastic line (i.e. $d_{x^{2}-y^{2}}$-related excitations).  We note in the inset of Fig. 2(a) our $d$-orbital convention relative to the crystal axes: the crystal $b$ axis is equivalent to the $z$ axis of the $d$-orbitals, and the $x/y$-axes of the $d$-orbitals lie along the copper-oxygen bond direction. Reproducing the experimental intensity variation leads to the assignment of $d$-orbital character for each of the excitations (annotated in Fig 2(b)):  $d_{xy}$ at 1.67 eV, degenerate $d_{xz/yz}$ at 1.99 eV, and $d_{3z^2-r^2}$ at 2.37 eV. The energies account for the geometric crystal field as well as ligand field hybridization effects within in the $d$-manifold. Given the geometry and local coordinates for each plaquette as well as the lack of confining apical oxygen or rock salt layers compared with two-dimensional planar cuprates, the assignment is consistent with the fact that both the $d_{xz/yz}$ and $d_{3z^2-r^2}$-orbitals possess more out-of-plane character and, therefore, should sit at higher energy loss compared to the $d_{xy}$-orbital that lies primarily in-plane.

In addition to the positions and intensities, we can also examine the lineshapes of the $dd$-excitations. Fig. 2(c) shows both Gaussian and Lorentzian fits to the data, focused near the edge of the first peak to emphasize the quality of fit. We find that the Gaussian fit gives a much smaller error compared with the Lorentzian fit, which is the expected lineshape for a single oscillator-like excitation. However, the width of the Gaussian peak is nearly twice that of our resolution, indicating that instrumental broadening is not the cause for such unusual lineshapes. Instead, such data strongly indicates the presence of additional underlying features.

To gain further insight, we perform an incident energy dependent study across the Cu $L_{\mathrm{3}}$-edge at specular reflection. The false-color RIXS map, shown in Fig. 3(a), highlights an unexpected shift in the $dd$-peak energies as one tunes incident photon energy across the resonance. To quantify this energy shift, we plot in Fig. 3(b) the energy positions of the $d_{xy}$ and $d_{xz/yz}$ peaks obtained from Gaussian fits to the data, fitting all three peaks simultaneously. We omit the $d_{3z^2-r^2}$ peak as the fitted peak position had much higher uncertainty due to lower spectral weight when moving away from the absorption edge. This graph reveals an unambiguous shift of ${\sim}50$-$80$ meV for the two plotted peak with increasing incident photon energy across the edge. We again note that the energy scale for magnetic exchange $J$ is too small to likely account for the size of this shift without invoking extremely large electron-spin coupling. Interestingly, previous RIXS experiments at the O $K$-edge\cite{PhysRevLett.110.265502} on the same compound identified significant electron-lattice coupling to an oxygen phonon mode at ${\sim}70$ meV in this system, similar to the observed energy shift for the two peaks. Furthermore, such energy scales are comparable to those seen in other cuprates through angle-resolved photoemission measurements, where ``kinks" in the electronic dispersion are signatures of electron-phonon coupling, with the 70 meV energy scale attributed to in-plane bond breathing modes\cite{BogdanovKink, CukReviewElPhCoupling}. Thus, it is likely that the observed energy shift is associated with coupling between the orbital $dd$-excitations and this 70 meV phonon.

We now extend the atomic model to include lattice degrees of freedom. However, as nuclear motion occurs on a much longer timescale than electronic excitation, we utilize the Born-Oppenheimer approximation in which the lattice and electronic degrees of freedom are decoupled and apply a Franck-Condon treatment of the interaction between the $dd$-excitations and phonons. A schematic of the Franck-Condon process applicable to RIXS at the Cu $L$-edge is shown in Fig. ~\ref{fig4}(c). The electronic potential energy of the system is determined by the lattice spacing between the copper and oxygen atoms, which can be parametrized by an effective internuclear coordinate, as sketched in Fig. 3(c), and whose equilibrium value is denoted by $\eta$. Starting in the ground state ($2p^{6}3d^{9}$) with equilibrium internuclear coordinate $\eta_{0}$, the first step in the RIXS process creates a $2p^{5}3d^{10}$ intermediate state. The additional electron in the valence shell increases the Coulomb repulsion with the ligands, causing an increase in the equilibrium coordinate by $\Delta\eta_{e}$. The photoexcitation occurs instantaneously compared to the time scale of nuclear motion, introducing a manifold of lattice vibrations into the intermediate state. In the second step of the RIXS process an electron is removed from the valence shell and a photon is emitted, leaving the system in an excited $2p^{6}3d^{9*}$ final state. This causes a decrease in $\eta$, related to the reduction in effective Coulomb repulsion with the ligands. This decrease, denoted $\Delta\eta^{\alpha}_{d}$, will depend upon the $d_{\alpha}$-orbital involved in the final state excitation. This relaxation is also instantaneous compared to nuclear motion, introducing another manifold of lattice vibrations into the final states. 

We allow the electronic contribution, i.e. the bare $dd$-energies and lifetimes, to be free fitting parameters. Such excitations will have a Lorentzian lineshape, with the energy and lifetime corresponding to the peak position and linewidth, respectively. The contribution from lattice coupling requires an estimate of the phonon energy ($\Omega$) and the phonon effective mass ($\mu$), as well as the change in effective internuclear coordinate during both the excitation ($\Delta\eta_{e}$) and de-excitation processes ($\Delta\eta^{\alpha}_{d}$)\cite{HancockNJP}. We fix the values of $\Omega=70$ meV and $\Delta\eta_{e}=0.22$ \AA~, in agreement with previous experiments that analyzed the influence of lattice coupling near the elastic line at the O $K$-edge\cite{PhysRevLett.110.265502}. We take the effective mass to be that of oxygen ($\mu=16$ amu). The $\Delta\eta^{\alpha}_{d}$'s are left as additional free fitting parameters, which control the overall energy shift of each $dd$-excitation. We plot the calculated RIXS energy loss map in Fig. ~\ref{fig4}(d) and apply the same analysis used on our experimental data, i.e. fitting the calculated spectra to three Gaussian peaks. The fitted positions for the d$_{xy}$ and d$_{xz/yz}$ peaks are plotted in Fig.~\ref{fig4}(e). The shifts seen in the experimental data for these peaks are well reproduced by the calculations. Effects of instrumental resolution were not included as they are secondary to the underlying cause of the peak lineshape distortion, discussed below.  

To further illustrate how the electron-phonon coupling manifests in the $dd$-excitation profile, we plot in Fig. 4 a representative energy loss curve (dotted curve), calculated for an incident photon energy of 931 eV. The contribution from individual phonon occupations (solid curves) are shown color-coded to each $dd$-excitation. The zero-phonon lines are plotted in bold and their respective positions, i.e. the bare $dd$-excitation energies, are marked with an arrow of corresponding color. As can be seen, the individual phonon lines overlap to produce a broadened $dd$-excitation peak with an effective Gaussian envelope (colored dashed curves), despite the fact that individual peaks have Lorentizan lineshape. This provides a natural explanation for the Gaussian-like $dd$-excitations observed in the experimental data (Fig. 2 (c)), beyond that of simple instrumental broadening.  Varying incident photon energy changes the relative weights of the individual phonon components, resulting in an apparent shift in the Gaussian envelope peak.  In addition, we notice a large difference between the peak positions obtained from the fitting compared to the peak positions of the bare excitations, which clearly demonstrates that the experimentally observed $dd$-excitations energies have been renormalized through electron-lattice coupling.

Similar FC-induced Gaussian lineshapes have been observed in the single particle spectral function in underdoped cuprates by angle-resolved photoemission spectroscopy\cite{PhysRevLett.93.267002}. In x-ray scattering studies, we note that shifts have been observed in charge-transfer excitations at the Cu $K$-edge\cite{HancockNJP, PhysRevLett.83.860, PhysRevB.69.155105}, and that in non-resonant inelastic x-ray scattering, anomalous peak widths of $dd$-excitations have been attributed to coupling to phonons\cite{IXS_dd_phonons}. However, in our study one can deduce the effects of the lattice coupling to electrons with different orbital character, as the $dd$-peaks are well-separated. Such an orbital-selective coupling could not be observed in measurements from the elastic or charge-transfer peaks.

\begin{figure}[t]
\includegraphics[scale = .35]{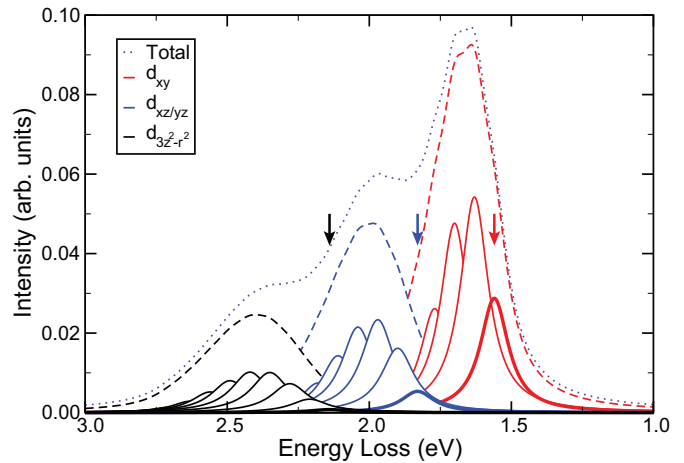}
\caption{Single RIXS spectrum calculated for an incident photon energy of 931 eV . The individual phonon peaks are plotted as solid lines, and color coded to their respective $dd$-excitation. The zero-phonon lines are plotted in bold, with the peak positions indicated with a corresponding colored arrow. The sum of the phonon peaks for each excitation are plotted as colored dashed lines, demonstrating an effective width broadening and emergent Gaussian lineshape. The total spectrum is plotted as a dotted black line, where the envelope peak positions (plotted previously in Fig. 3(e)) differ from the bare excitation positions. The energy difference is determined by the number of phonons which are excited, which depends on the change in equilibrium lattice constant.}
\end{figure}

Our results demonstrate that the coupling between low and high energy excitations, such as the charge-orbital-lattice coupling observed in our study, can manifest in the RIXS spectrum. We also emphasize that this coupling will affect the electronic parameters derived from spectroscopic measurements of the associated excitations. For example, the derived ligand-field energy-splitting of the $dd$-excitations will be larger than the ``true" values as the measured peak positions are shifted to higher energy loss due to this coupling with lattice excitations. In addition, the derived excitation lifetimes will be underestimated as the peak linewidths are broadened due to the overlap of multiple phonon excitations, beyond that of instrumental resolution. With additional improvements in resolution and theoretical treatments of RIXS, detailed studies such as the one presented here will not only help identify the constituent excitations of solids but also help elucidate the interactions between them.

The authors would like to thank S. Johnston and J. van den Brink for valuable discussions. This work was supported by the U.S. Department of Energy, Office of Basic Energy Science, Division of Materials Sciences and Engineering under Contract No. DE-AC02-76SF00515 and through the CMCSN program under Grant No.~DE-SC0007091. J.J.L. acknowledges funding from the Stanford Graduate Fellowship Program. The experiments were performed at the ADRESS beam line of the Swiss Light Source at the Paul Scherrer Institut.

\bibliography{RIXS_dd_Bib}

\end{document}